\documentclass[
reprint,
superscriptaddress,
preprintnumbers,
 amsmath,amssymb,
pra
floatfix]{revtex4-2}

\usepackage{braket}
\usepackage{comment}
\usepackage{graphicx}
\usepackage{float}
\usepackage{tikz}
\usepackage{dcolumn}
\usepackage{bm}
\usepackage{hyperref}
\usepackage{mathtools}
\usepackage [english]{babel}
\usepackage [autostyle, english = american]{csquotes}
\usepackage{comment}
\MakeOuterQuote{"}

\begin{document}


\title{Quantum Imaging with X-rays}


\author{Justin C. Goodrich}
\email[]{jgoodrich@bnl.gov}
\affiliation{National Synchrotron Light Source II, Brookhaven National Laboratory, Upton, NY 11973}

\author{Ryan Mahon}
\affiliation{National Synchrotron Light Source II, Brookhaven National Laboratory, Upton, NY 11973}

\author{Joseph Hanrahan}
\affiliation{National Synchrotron Light Source II, Brookhaven National Laboratory, Upton, NY 11973}

\author{Dennis Bollweg}
\affiliation{Computing and Data Sciences Directorate, Brookhaven National Laboratory, Upton, NY 11973}

\author{Monika Dziubelski}
\affiliation{Physics Department, Brookhaven National Laboratory, Upton, NY 11973}

\author{Raphael A. Abrahao}
\affiliation{Physics Department, Brookhaven National Laboratory, Upton, NY 11973}

\author{Sanjit Karmakar}
\affiliation{Department of Physics and Astronomy, Stony Brook University, Stony Brook, NY 11790}

\author{Kazimierz J. Gofron}
\affiliation{Spallation Neutron Source, Oak Ridge National Laboratory, Oak Ridge, TN 37831}

\author{Thomas A. Caswell}
\affiliation{National Synchrotron Light Source II, Brookhaven National Laboratory, Upton, NY 11973}

\author{Daniel Allan}
\affiliation{National Synchrotron Light Source II, Brookhaven National Laboratory, Upton, NY 11973}

\author{Lonny Berman}
\affiliation{National Synchrotron Light Source II, Brookhaven National Laboratory, Upton, NY 11973}

\author{Andrei Nomerotski}
\affiliation{Department of Electrical and Computer Engineering, Florida International University, Miami, FL 33174}
\affiliation{Faculty of Nuclear Sciences and Physical Engineering, Czech Technical University in Prague, 11519 Czech Republic}

\author{Andrei Fluerasu}
\affiliation{National Synchrotron Light Source II, Brookhaven National Laboratory, Upton, NY 11973}

\author{Cinzia DaVià}
\affiliation{Department of Physics and Astronomy, Stony Brook University, Stony Brook, NY 11790}
\affiliation{Department of Physics and Astronomy, The University of Manchester, M13 9PL, Manchester, UK}

\author{Sean McSweeney}
\email[]{smcsweeney@bnl.gov}
\affiliation{National Synchrotron Light Source II, Brookhaven National Laboratory, Upton, NY 11973}
\affiliation{Biology Department, Brookhaven National Laboratory, Upton, NY 11973}


\date{\today}

\begin{abstract}
Quantum imaging encompasses a broad range of methods that exploit the quantum properties of light to capture information about an object. One such approach involves using a two-photon quantum state, where only one photon interacts with the object being imaged while its entangled partner carries spatial or temporal information. To implement this technique, it is necessary to generate specific quantum states of light and detect photons at the single-photon level. While this method has been successfully demonstrated in the visible electromagnetic spectrum, extending it to X-rays has faced significant challenges due to the difficulties in producing a sufficient rate of X-ray photon pairs and detecting them with adequate resolution. Here, we demonstrate record high rates of correlated X-ray photon pairs produced via a spontaneous parametric down-conversion process and we employ these photons to perform quantum correlation imaging of several objects, including a biological sample (E. cardamomum seedpod). Notably, we report an unprecedented detection rate of about 6,300 pairs per hour and the observation of energy anti-correlation for the X-ray photon pairs. We also present a detailed analysis of the properties of the down-converted X-ray photons, as well as a comprehensive study of the correlation imaging formation, including a study of distortions and corrections. These results mark a substantial advancement in X-ray quantum imaging, expanding the possibilities of X-ray quantum optical technologies, and illustrating the pathway towards enhancing biological imaging with reduced radiation doses.
\end{abstract}


\maketitle

\section{Introduction}
The ability to visualize systems with high accuracy is crucial for effective analysis and understanding. However, high accuracy measurements require the production of images with high signal-to-noise ratios. Typically, this is achieved using high input flux, but for living cells a high incident X-ray dose complicates the image by inducing unwanted artifacts from radiation damage. This issue represents the essential compromise made in designing a biological imaging experiment: the experimenter must choose between precision of the image and damage to the sample. New experimental methods to allow for low-dose, high precision measurements are urgently needed.

The generation, manipulation, and detection of single photons have driven numerous advancements in quantum information science and technology. A key method for producing correlated single photon pairs is spontaneous parametric down-conversion (SPDC)\cite{SPDC_general,mosley2008heralded,grice2001eliminating,Brianna2023_SPDC}. SPDC is a technique which has been transformative in both fundamental quantum physics\cite{Kwiat,LoopholefreeBell_Vienna_TES,Loopholefree_Bell_NIST_nanowire,OAMentanglement} and various quantum technologies~\cite{dowling&milburn,quantum_cnot,quantum_micros_HOM,moreau2019qimaging,Pittman,harder2013optimized}. During SPDC, a photon from a pump beam is down-converted in a non-linear material into a pair of correlated photons. This process is widely applied in the visible and infrared spectral regimes. However, adapting this method to X-ray energies poses significant challenges, including the low conversion efficiency of commonly used media, such as single crystal diamond, and the absence of rapid, energy-resolving X-ray detectors.~\cite{Freund2,Eisenberger,Freund,Yoda,Adams,Adams2001,adams2013x_review,Sofer2019,Li2017,Schori2017,Schori2018-xv,Shwartz,apsspdc,Borodin,wong2021prospects}.

In this study, we present advances in the understanding of both the theoretical and the experimental aspects of X-ray SPDC. We have successfully performed quantum correlation imaging of various objects, including a complex biological sample (an E. cardamomum seed), with X-ray SPDC photons detected at a rate approximately 20 times higher than previously reported values~\cite{Borodin}. Additionally, we explore the effects of the down-converting medium crystalline quality along with the beam divergence on the properties of the biphotons, and discuss the pathway toward achieving quantum-enhanced sample transmission measurements via the so-called sub-shot-noise imaging modality. These achievements lay the foundations for the use of entangled X-rays to image radiation-sensitive biomaterials with a reduced dose~\cite{BioDamage,BioDamage2}. Our findings, along with initial work from other groups~\cite{Sofer2019}, underscore the potential of X-ray quantum and ghost imaging to surpass classical methods in resolution and signal-to-noise ratio by leveraging quantum principles~\cite{wong2021prospects,QuantumIllum,QuantumIllum2,QuantumIllum3,moreau2019qimaging,rayleigh_limit,noise_reduction,sub_shot_noise}, as well as providing a new regime for the study of fundamental quantum physics.

\section{X-ray Quantum Correlation Imaging}

\begin{figure*}
\includegraphics[width=1\textwidth]{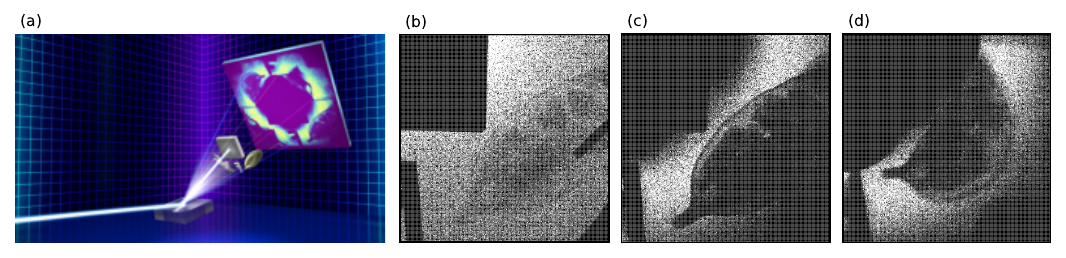}
\caption{Quantum X-ray correlation imaging. (a) A conceptual schematic of the non-linear X-ray diffraction (SPDC) imaging setup. 15 keV pump X-rays from a synchrotron light source produce a ring of SPDC biphotons from the non-linear diffraction of a diamond crystal. Test objects made of tungsten, shaped like a cat and the letter `F', along with an E. cardamomum seed, are placed within the ring on the lower two chips of the detector, while the upper chips remain unobstructed. Coincidence measurements yield a quantum direct (signal) correlation image of the objects on the lower chips and a deformed quantum ghost (idler) image appearing on the upper chips. (b) Reference image of the cardamom seed captured using a classical light source of scattered 15 keV pump X-rays. (c) The quantum direct correlation image of the seed on the signal detector with photon energies probing the samples ranging from 5 to 9 keV. (d) The quantum ghost correlation image of the seed on the idler detector, rotated to align with the signal detector's frame of reference for easier comparison, showing distortion due to the non-degeneracy of the X-ray photon pairs. Figures (b)-(d) all show the same number of photons per plot.}
\label{fig1}
\end{figure*}

Quantum imaging offers significant advantages over classical imaging across several metrics~~\cite{quantum_microscope,moreau2019qimaging,Lemos2014}. These include super-resolution capabilities that surpass the Rayleigh diffraction limit~\cite{rayleigh_limit}, noise reduction by measuring signals only coincident in two independent detection systems~\cite{noise_reduction}, and enhanced signal-to-noise ratio (SNR) in transmission imaging under ultra-low light levels through correlating the so-called signal and idler photon coincidence counts (sub-shot-noise imaging)~\cite{sub_shot_noise,sub_shot_noise2}. Despite significant advances in the optical regime, to date these techniques have remained largely inaccessible in the X-ray regime due to the aforementioned challenges in the generation and detection of sufficient numbers of correlated X-ray pairs.

In our proof-of-concept experiment of imaging complex biological samples, we utilize the simultaneous detection of X-ray biphotons with a high-speed, two-dimensional pixelated area detector. This setup enables the production of quantum correlation radiographs of various objects, including the intricate biological structure of an E. cardamomum seedpod. To our knowledge, this is the first quantum correlation X-ray imaging experiment performed on a biological object with a complex, low density internal structure.

As illustrated in the conceptual schematic describing the experiment (Fig. \ref{fig1}(a)), the characteristic ring generated by the correlated photon-pairs is recorded by a Lynx T3 pixelated area detector, which consists of four Timepix3 silicon ASICs. The top two chips are used to detect idler photons, unobstructed by the objects under study, while the bottom two chips detect signal photons on the objects' paths. By correlating coincident photon events in both detector arms, we generate two distinct quantum images, each from the respective frame of reference of the detectors (signal and idler, respectively). Fig. \ref{fig1}(b)-(d) show images of the cardamom seed using a classical source (scattered pump X-rays from the diamond crystal), as well as the quantum images with comparable photon counts. The image distortion (Fig. \ref{fig1}(d)) is due to the presence of non-degenerate pairs, which distribute on the ring non-linearly to satisfy conservation of momentum. More information on this process will be discussed in detail in the \textit{SPDC Photon Energy and Spatial Properties} section. The use of quantum correlations has the inherent property of reducing detector noise, as any electronic noise or cosmic rays incident on the sensor would have to exist simultaneously in both signal and idler arms to be marked as coincident. On the other hand, other sources of noise, such as that from scattered background, are always present due to their arrival time being within the time resolution of the detector and therefore require a more sophisticated pair-selection process. Improved detection capability of non-correlated single photons will allow for the reduction of uncertainty of the number of incident photons for projection imaging, useful for roentgenography and computed tomography measurements with significantly reduced dose~\cite{sub_shot_noise}, as will be discussed in the \textit{Towards Sub-Shot-Noise Transmission Imaging} section.

\section{Experimental procedure and methodology}

Our investigations were conducted at the 11-ID Coherent Hard X-ray Scattering (CHX) beamline of the National Synchrotron Light Source II (NSLS-II) facility at Brookhaven National Laboratory (BNL). The beamline was configured to produce a monochromatic X-ray beam set at 15 keV. The average input flux of the pump beam generated was approximately $10^{11}$ photons/second, with a polarization of 99\% in the horizontal direction and $\frac{\Delta E}{E} \approx 10^{-4}$. The dimensions of the 15 keV input beam were set to 50 $\mu$m (horizontal) x 50 $\mu$m (vertical). The incident X-ray beam was directed onto a (111) diamond single crystal with dimensions of 3 mm x 3 mm x 0.33 mm, grown by chemical vapor deposition from Element Six~\cite{element6}, which was used as the non-linear medium for the generation of correlated photon pairs. A vacuum-pumped flight path beam pipe was installed between the diamond crystal and the detector surface. The total path distance from the crystal to the detector surface was 68.3 cm. This distance was confirmed from the scattering pattern of a standard SAXS distance calibration sample, silver behenate.

After standard height alignment procedures, the diamond crystal was oriented to the (111) Bragg reflection at a Bragg ($\theta$) angle of $11.576^\circ$. Multiple positions along the surface of the diamond were probed to find the area with the highest quality, using rocking curves ($\theta$ scans) to minimize dual peaks and full-width-at-half-max (FWHM). The resulting diffraction pattern peaked around a pixel coordinate of (260, 256) on a Lynx T3 (Timepix3) detector by Amsterdam Scientific Instruments, which was subsequently employed for measurement of the SPDC signal.

\begin{figure}
\includegraphics[width=1.0\columnwidth]{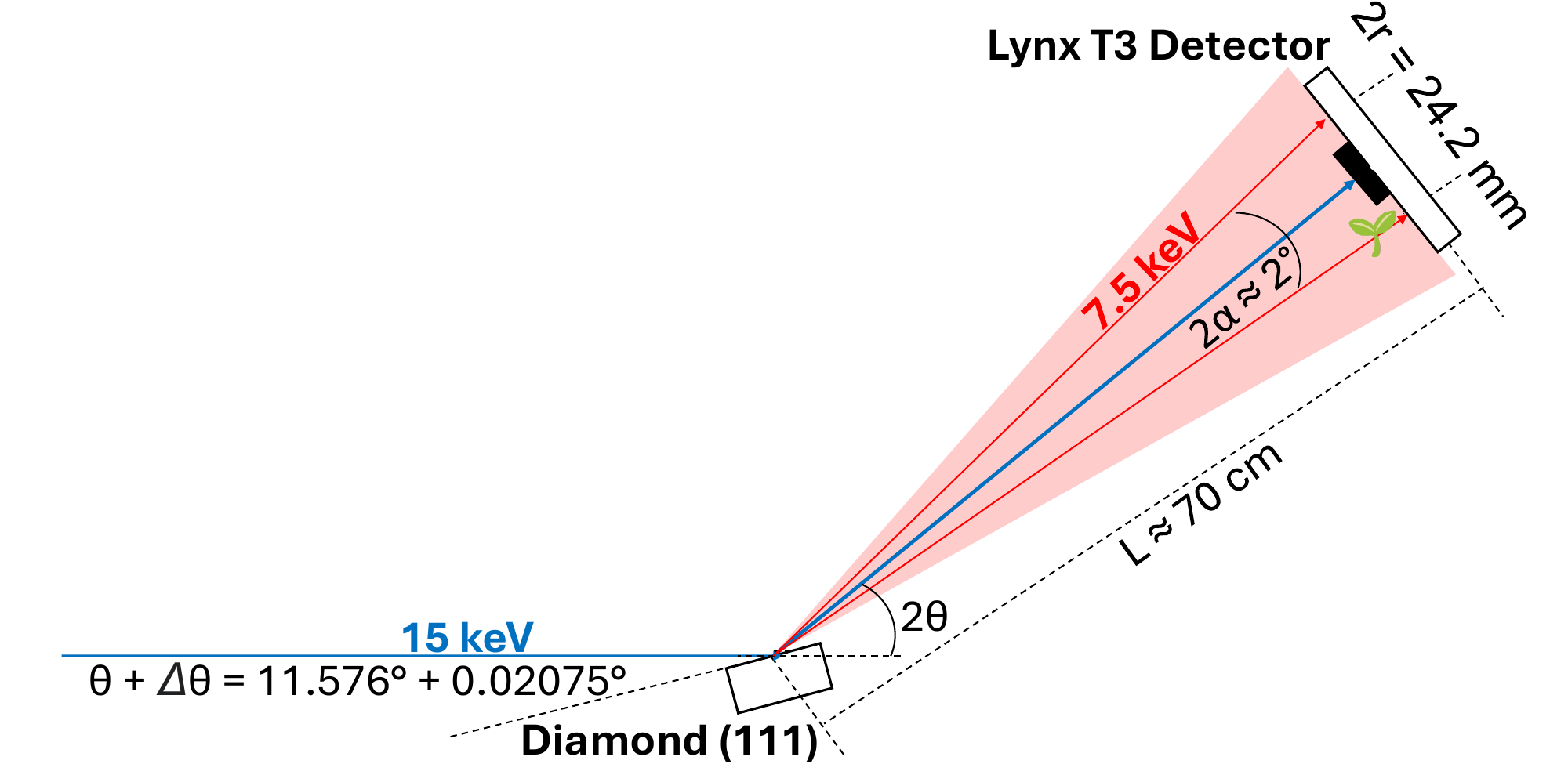}
\caption{Detailed Experimental Setup. Detailed X-ray SPDC experimental setup indicating the geometry and positions of the detectors (not to scale). A partially coherent 15 keV pump from the CHX beamline is incident upon a diamond crystal slightly detuned ($\sim$21 millidegrees) from the (111) Bragg condition. This causes satisfaction of the non-linear diffraction phase-matching condition, creating a diffuse ring of down-converted single photon pairs. The Lynx T3 detector is placed approximately $\sim$70 cm away at the 2$\theta$ angle to measure such photons, with a tungsten beamstop positioned to block the direct Bragg reflection. Imaging samples are placed on one side of the detector for the coincidence imaging experiments.}
\label{efig1}
\end{figure}

A tungsten beamstop was strategically positioned to obscure the direct Bragg reflection and nearby scattering. The Bragg alignment was then fine-tuned with a deviation of $\sim$0.021$^\circ$ to a $\theta$ angle of 11.598$^\circ$ to meet the phase matching condition. This adjustment instantiated the production of correlated X-ray pairs at a peak output angle of $\sim$1.00$^\circ$, an angle chosen to maximize the SPDC signal between the edge of the detector and the tungsten beamstop with the detector placed 68.3 cm away from the diamond at the 2$\theta$ position (Fig. \ref{efig1}). 

To isolate the correlated X-ray photon pairs, we employed the Lynx T3 detector~\cite{UsingTimePix}. This detector combines four 256 x 256 pixel arrays of pixel size 55 x 55 $\mu$m, resulting in a total detector area of 512 x 512 pixels~\cite{timepix3}. The top two chips were isolated for use as ``idler" detector whereas the bottom two chips were considered to be the ``signal" detector for the quantum ghost imaging experiments. This detector was chosen for its accuracy in measuring the time-of-arrival (ToA) and energy deposition (time-over-threshold or ToT). Two photons generated in an SPDC process should be coincident in time when reaching the detector. 
However, the signal amplitude dependence in the timestamping process results in a "timewalk"~\cite{Timewalk, Timewalk2}, whereby events that arrive at the same time might measure a time difference of up to 100 ns. The time resolution could be further deteriorated by different time delays of ionization drift in silicon. This is due to the variation of X-ray conversion points since the absorption length, about 100 um, is of the same order as the sensor thickness, 300 um. The measured resolution (rms) for the two x-rays time difference was determined to be equal to 18 ns.
The detector was integrated into the NSLS-II CHX beamline's control system through the development of an appropriate EPICS areaDetector driver~\cite{AD,ADTimepix,ADTimepix2} and Python objects for integration with the Bluesky RunEngine~\cite{Bluesky}.

A minor fraction of hot pixels ($\sim$0.1\%) were masked to avoid recording substantial amounts of noise hits. This system allows for multihit functionality for each pixel, independent of others, along with a rapid readout bandwidth of 120 MPix/sec (30 MPix/sec/chip). Occasionally, photons excite charge in multiple adjacent pixels; to correct for this, a k-d tree algorithm was employed to cluster and centroid such events together. The 512 x 512 pixel array has two columns and two rows along the center (x and y = 255 and 256) of large, double-length (110 x 55 $\mu$m) interpixels. This is corrected by inserting rows and columns of two dummy pixels between chips to preserve the correct physical distance between pixels.   

We calibrated the ToT dependence on X-ray energy by allowing the detection of beams of monochromatic scattered X-rays, varying their energy from 6 to 15 keV in 0.5 keV steps. We also cross-calibrated the ToT energy estimator using the spatial information of selected SPDC pairs, which gave good agreement from the scattering calibrations. These two calibration techniques provide reliable and redundant ways to determine the X-ray energy, with the second approach only applicable for down-converted X-rays, but providing considerably more precise energy determination. The dual-calibration method ensures robust accuracy, leveraging spatial information as a complementary validation to the ToT measurements. These calibration studies identified pixel-by-pixel variations in the ToT response as a function of energy, which allowed for the selection of cutoff values for the isolation of the SPDC pairs on an individual pixel basis. These cutoff values were employed to reduce the high background of scattered 15 keV pump photons from the hits of lower energy (and thus lower ToT) SPDC single photons.

The objects, including the tungsten cat, F, and E. cardamomum seed, studied in the quantum ghost imaging setup were attached to a Kapton window via a quick drying epoxy. They were positioned in such a way to occlude the incident X-rays towards two adjacent of the Timepix3 ASIC chips, which were considered to be the signal photon detectors. The other two ASICs were left obstructred and considered as idler photon detectors. The data presented in Fig. \ref{fig1} and Fig. \ref{fig4} represent 38 hours of collection. 

All data processing and analysis were conducted using custom Python code, developed in-house and computed on CHX's local resources. This tailored approach facilitated the meticulous processing of the raw Lynx T3 detector output, including hit clustering, distance corrections, data filtering, time coincidence searches, photon pair determinations, and analysis of spatial and energy correlations. The simulations were also performed using Python.

\section{Generation and Identification of SPDC X-rays}

\begin{figure}
\includegraphics[width=1.0\columnwidth]{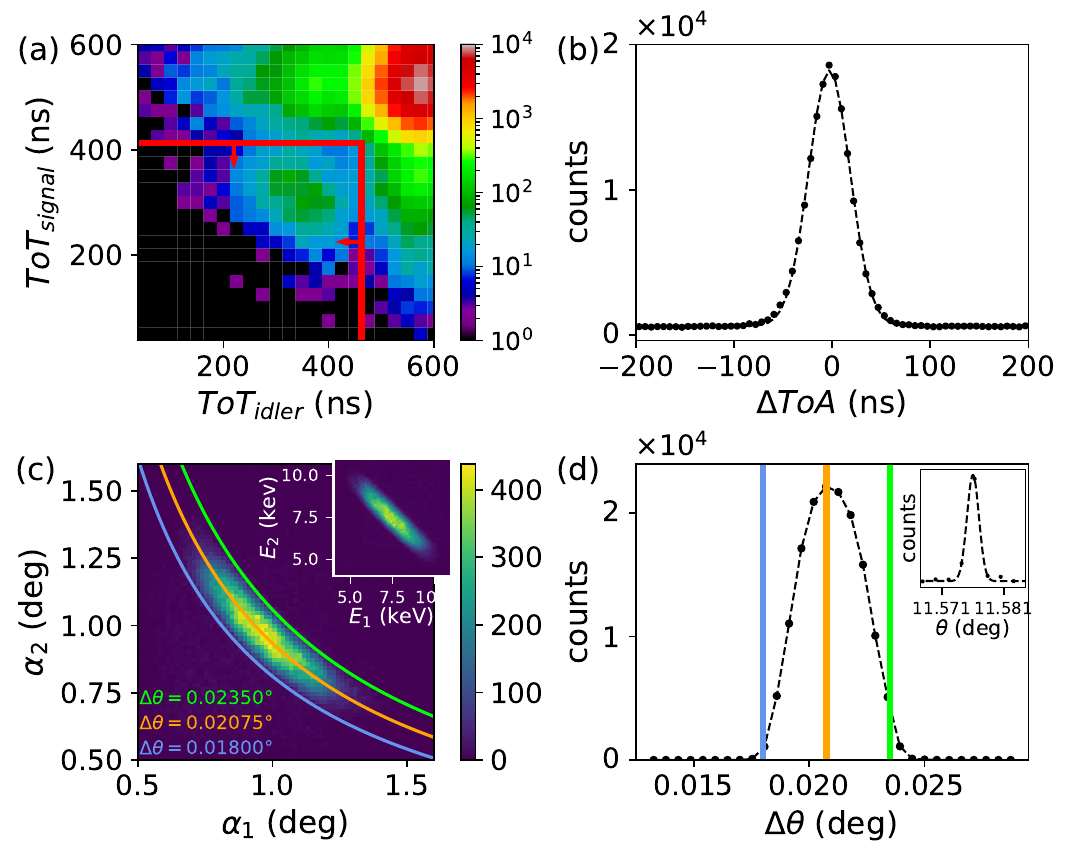}
\caption{X-ray SPDC selection observables. (a) 2D histogram of photon pair ToTs when events from the test and reference detectors are paired by smallest time difference. This dataset is from a 20 minute exposure. The peak at smaller ToT coordinates (area indicated by red lines) represents SPDC photon pairs, in contrast to background pump photons at large ToT coordinates. These events in the lower left section, indicated by the red arrows, are selected for further analysis. (b) A pronounced time coincidence signal emerges after ToT and spatial filtering, featuring a narrow width due to the timewalk phenomenon, which confirms the simultaneous measurement of the SPDC photons isolated from scattered pump background events. This dataset is from 38 hours of exposure with the imaging mask obstructing the detector. (c) Calculated emission angles and photon energies (inset) for the isolated SPDC single photons correlate well with theoretical predictions based on SPDC phase-matching conditions, affirming energy conservation and indicating the process's non-degeneracy. This dataset is from 38 hours of exposure with the imaging mask obstructing the detector. (d) The calculated detuning angles of the pairs, based on their emission angles, show a spread of detuning angles around the nominal (set) detuning angle of 0.021$^\circ$. The inset displays the measured diamond (111) rocking curve on the same x-axis scale, providing observational evidence that the spread in detuning angles is correlated to the crystalline misorientation, beam divergence, and Darwin width indicated by $\theta$ scans. This dataset is from 38 hours of exposure with the imaging mask obstructing the detector.}
\label{fig2}
\end{figure}

Phase-matching for the down-conversion occurs when the crystal is detuned very slightly from the Bragg condition~\cite{Freund,Freund2,Eisenberger,Adams2001}. The extremely low conversion efficiency of X-ray SPDC, which is less than $10^{-10}$, significantly limits the number of generated photon pairs in comparison to similar processes in the optical regime. This difficulty is exacerbated by a substantial background of scattered pump photons that also reach the detector, masking the signal. However, the application of Timepix technology, which records both the event time-of-arrival (ToA) with $\sim$1.56 ns precision, and time-over-threshold (ToT) with 25 ns precision, facilitates the pairing of photons based on the smallest time differences after filtering out high-energy background events (Fig. \ref{fig2}(a)). Selecting such events allows for the detection of photon pairs with strongly correlated emission angles emerging from the diamond crystal and a robust time coincidence signal centered at a time difference of 0 ns. Although the time precision of the Timepix3 chip is 1.5625 ns, the aforementioned ``timewalk'' phenonemon, along with variations in the X-ray conversion points in the silicon detector, result in an effective time resolution on the order of 20 ns (Fig. \ref{fig2}(b)). Further refinement of the selected events based upon spatial properties, such as pairs that are seen to conserve both the diffracted pump energy and momentum, allow for the isolation of SPDC photon pairs with high ($>$100).

In our experiments, the detuning angle was set to 0.021$^\circ$, an angle that was chosen to position the half-energy ring on the detector between the detector edges and the tungsten beamstop which blocks the direct Bragg reflection. Despite the detuning angle being locked in such a way, the SPDC photon pairs are measured to have a spread of emission angles $\alpha_s$ and $\alpha_i$ on the order of 0.1$^\circ$ (Fig. \ref{fig2}(c)). It is possible to compute the value of the detuning angle from each measured pair by applying the energy conservation law and knowing the existing inverse relationship between energy and emission angle for each photon (see Appendix 2 for derivation):

\begin{equation}
\Delta \theta = \frac{\alpha_s \alpha_i}{2 \sin(2 \theta)}
\end{equation}

This calculation suggests that the SPDC photons are generated with a spread of detuning angles, centered around the preset angle with a standard deviation ($\sigma$) of 0.0014$^\circ$ (Fig. \ref{fig2}(d)). The observed variance in the detuning angle corresponds to the variance of the measured diamond (111) Bragg peak rocking curve during the crystal alignment. For reference, the Darwin width of the diamond(111) Bragg reflection at 15 keV is 0.75 mdeg, and the vertical divergence of the incident beam, taking account of the source and slit sizes and distance separating them, was 0.1 mdeg. Comparing with the measured rocking curve FWHM of 3 mdeg, the crystal mosaicity is the predominant contribution to the rocking curve width. This correlation provides strong evidence that the mosaicity of the crystal, along with any beam divergence, influences the detuning of the SPDC process in a way comparable to the effect of such influences when performing $\theta$ scans (rocking curves). Furthermore, it indicates the importance of minimizing beam divergence and using a high quality non-linear medium for the generation of robust X-ray photon pairs; achieving the latter, for a diamond crystal, entails using as perfect a crystal as possible with a minimum of mosaicity.

The final selections indicate SPDC rates approaching $6.3 \times 10^3$ pairs per hour, significantly surpassing previous benchmarks reported in the literature~\cite{Borodin}. This is mostly attributed to the improved detector performance, along with the high brightness of the CHX beamline. Detecting pairs at this rate with a two-dimensional pixelated area detector (Fig. \ref{fig3}(b)) substantiates the practicality of utilizing the X-ray SPDC source for quantum imaging experiments, which were previously limited due to low count rates and the utilization of slit-based detectors~\cite{Schori2018-xv}. Additionally, identifying the spread of detuning angles sheds light on the influence of pump and diamond properties on the generation of SPDC biphotons. This understanding facilitates using sharpening corrections for the image blurring which arises from the pair emission angle spread. Such corrections can enhance image clarity and precision while providing important insights on the best pump characteristics, crystalline quality, and the detuning angles necessary for the production of SPDC pairs with minimal emission angle dispersion.

\section{SPDC Photon Energy and Spatial Properties}

\begin{figure}
\includegraphics[width=1.0\columnwidth]{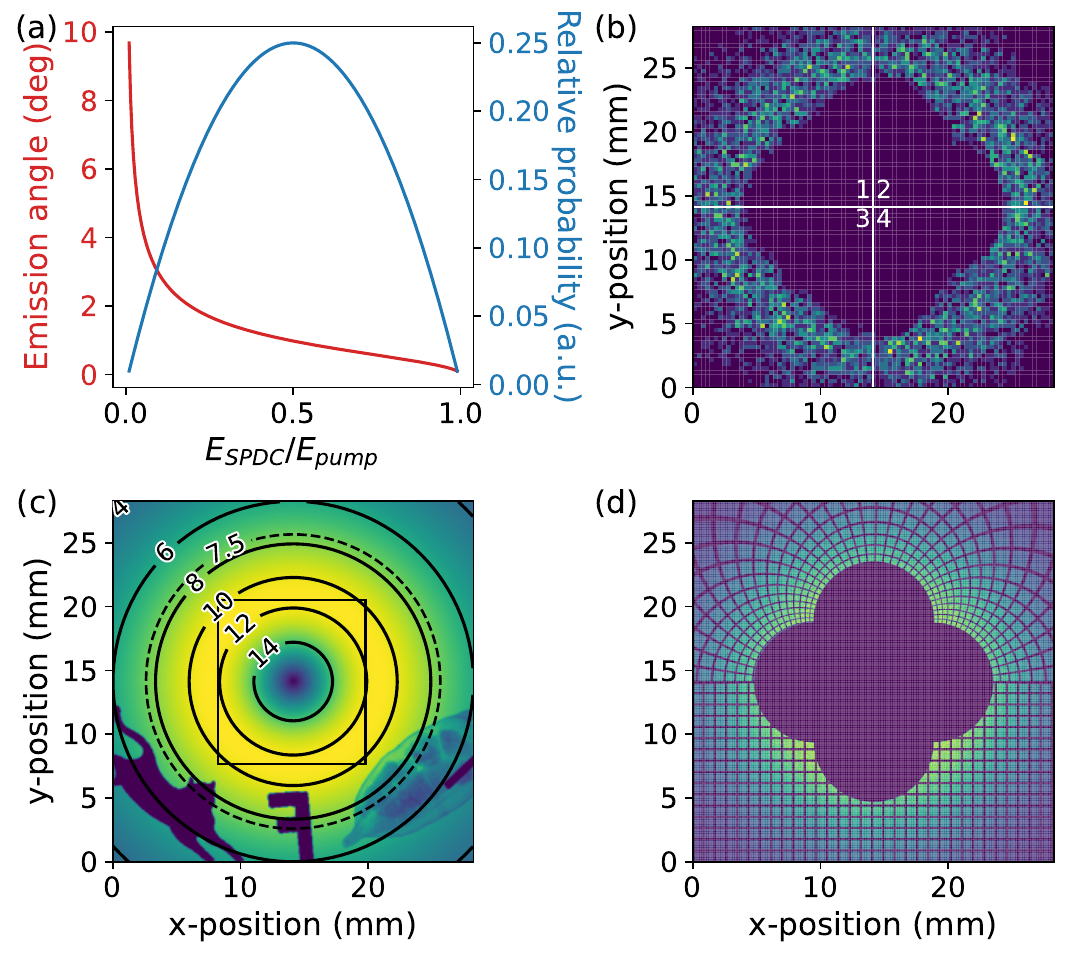}
\caption{X-ray SPDC Properties. (a) Radial distances and relative probabilities of the X-ray SPDC photons as a function of angle. (b) Experimentally-measured positions of the correlated down-converted pairs, distributed in the characteristic ring pattern, from a one exposure (no samples in the SPDC ring). The dark region in the center is where coincidences are lost because a corresponding photon in the pair is outside the detector area on the opposite side. The full detector consists of four detector silicon ASICs, labeled one through four. In subsequent ghost imaging experiments, the ghost image appears on ASICs one and two from objects placed in front of ASICs three and four. (c) SPDC X-ray energy contours and occupation density with a quantum correlation imaging mask of the tungsten cat, letter `F', and E. cardamomum seedpod. (d) Illustration of how a grid on the signal detector (bottom) maps onto circular contours on the idler detector (top) due to energy non-degeneracy.}
\label{fig3}
\end{figure}

A full understanding of the energy and spatial properties of the X-ray SPDC source is essential for its successful application in quantum imaging. The SPDC process generates correlated single photons over a broad energy spectrum given by the probability:

\begin{equation}
\eta(E_{SPDC}) \propto E_{SPDC}(E_{pump}-E_{SPDC})
\end{equation}

where $E_{SPDC}$ is the energy of an SPDC photon and $E_{pump}$ is the energy of the X-ray pump. The highest probability of generating an SPDC pair happens at degeneracy, where $E_{signal} = E_{idler} = \frac{1}{2} E_{pump}$. However, a significant number of photons are also generated in a non-degenerate condition. The angle between an SPDC photon's wave vector $\textbf{k}_{SPDC}$ and the diffracted pump $\textbf{k}_{out}$ is given by a function $\alpha(b)$, where b = $\frac{E_{SPDC}}{E_{pump}}$. The distance \text{r} between the SPDC photon and the diffracted pump on the detector is given by:

\begin{equation}
r = L \tan(\alpha(b))    
\end{equation}

where L is the distance of the detector from the diamond crystal. Considering the phase-matching condition and conservation of momentum, with some approximations (see Appendix 1) one can relate the energy of a photon with its position by using the following equation:

\begin{equation}
    E_{SPDC}(r) \approx \frac{E_{pump}}{\frac{\arctan^2(\frac{r}{L})}{2 \Delta \theta sin(2 \theta)} + 1}
\end{equation}

The expected photon occupation and spatial distances as a function of the single photon energy fraction $b$ can be determined by combining equation (4) and equation (2) (Fig. \ref{fig3}(a)). The detected degenerate and non-degenerate photon pairs distribute on precisely defined energy rings, allowing the sample under study to be exposed at different X-ray energies over its cross sectional area, depending on the radial distance (Fig. \ref{fig3}(b),(c)). An SPDC single photon detected at position $(x,y)$ with a radial distance $r$ from the diffraction center will have its idler counterpart incident at a position $(x',y')$ with radial distance $r'$ given by:

\begin{align}
r' \approx L \tan{\left( \frac{2 \Delta \theta \sin(2 \theta)}{\arctan{( \frac{r}{L} )} } 
 \right)} \approx \frac{\tan{\left(\sqrt{2 \Delta \theta \sin(2 \theta) \frac{b}{1-b}}\right)}}{\tan{\left(\sqrt{2 \Delta \theta \sin(2 \theta) \frac{1-b}{b}}\right)}}
\end{align}

The effect of this mapping is demonstrated in Fig. \ref{fig3}(d), where a square grid on the signal detector maps onto circular contour ghost images on the idler detector.

The non-degenerate properties of the source indicate the possibility of engineering a system to perform lensless geometric magnification by correlating high energy X-rays probing a sample with their lower energy counterparts at steeper emission angles. It also suggests the capability of an imaging mechanism where the detuning angle is swept across a range of angles, exposing the samples to different energies over time, allowing for quantum X-ray absorption and fluorescence spectroscopy schemes. 

\section{Spatial Correlations and Detuning Corrections}

\begin{figure}
\includegraphics[width=1.0\columnwidth]{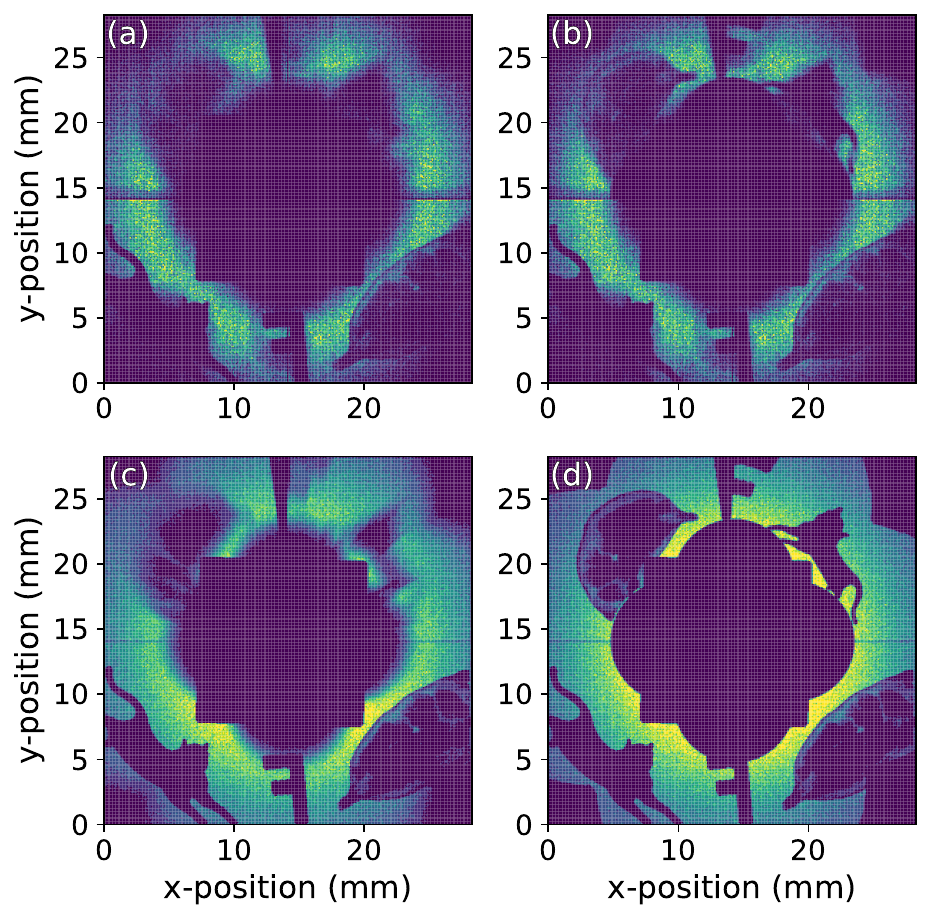}
\caption{X-ray correlation imaging simulations and experimental data. (a) Experimental X-ray correlation image before correction (38 hour exposure). (b) Experimental data after applying corrections from Equation (6) to photons on the top (idler) detectors. (c) Simulated X-ray correlation image with the SPDC detuning angles following a normal distribution centered at 0.021$^\circ$ with a standard deviation ($\sigma$) of 0.0014$^\circ$. (d) Simulated correlation image with the SPDC detuning angle fixed at 0.021$^\circ$.}
\label{fig4}
\end{figure}

One key property of the X-ray quantum signal and idler correlation images (Fig. \ref{fig4}(a)) is their dual mirroring and flipping relationship. This can be most evidently observed with the cat, whose head is positioned on the inner side of the ring on the signal detector (lower left quadrant), appearing on the outer part of the ring in the idler detector (upper right quadrant). This is also apparent with the letter `F', where the top horizontal section of the letter (which grazes against the semi-circular `dead area') in the lower direct image is imaged on the top of the upper detector. Similarly, the cardamom seed projects a direct image on outer ring radii which are mirrored and flipped on the idler detector onto inner radii near the beam stop. This effect is due to the correlation of low emission angle, higher energy X-rays being absorbed by the objects being correlated with steeper emission angle, lower energy X-rays. Also apparent in all three images is the mapping of straight segments into circular contours due to the energy non-degeneracy. The effect of this transformation becomes more extreme the further away the object is positioned from the degenerate energy ring.

Furthermore, an aberration is present in the quantum ghost correlation images due to the biphotons being generated with a spread of detuning angles. This points to the necessity of minimizing the pump divergence, to using a high quality single crystal medium, and to maximizing the detuning angle in order to reduce the impact of $\frac{\sigma_{\theta}}{\Delta \theta}$ - the ratio of the rocking curve width ($\sigma_{\theta}$) with the detuning angle ($\Delta \theta$). However, increasing the detuning angle leads to other trade-offs; the emission angles of the biphotons increase, which requires either a larger detector while maintaining detector distance or moving the detector closer to the diamond, thus sacrificing the effective resolution.

One possible approach to correct this blurring is to scale the radial distances of events on the idler detector to distances which would have resulted from the nominal detuning angle (Fig. \ref{fig4}B) via a form factor expressed by the ratio $\frac{\theta_{nominal}}{\theta_{calculated}}$:

\begin{equation}
r'_{corrected} = L \tan\left(\frac{\Delta \theta_{nominal}}{\Delta \theta_{calculated}} \alpha'\right)
\end{equation}

Simulations (Fig. \ref{fig4}(c),(d)) of the detuning angle spread match well with the experimental data and verify the utility of the correction equation (6). The SPDC correlation images vividly illustrate the distinct ring pattern characteristic of the spatial correlations between X-ray photon pairs. This pattern is a direct manifestation of the quantum mechanical nature of the photon pairs and also a validation of the theoretical predictions of the SPDC X-ray properties. The ring pattern observed includes the regions where the correlation is lost due to one of the single photons being off the plane of the detector or occluded by the beamstop. This reveals the importance of control over the experimental parameters (detector size, distance to detector, and detuning angle) in efficient photon pair detection. The different image quality between the corrected experimental and simulated data indicate the presence of other noise sources which could not escape the limited time and the spatial resolution of the detection system. 

\section{Towards Sub-shot-noise Transmission Imaging}

The ultimate goal of the quantum imaging scheme presented herein is to demonstrate sub-shot-noise imaging. Sub-shot-noise imaging refers to reducing noise below the shot-noise limit, which is determined by the Poissonian statistics of classical light sources \cite{sub_shot_noise,sub_shot_noise2}. In traditional transmission imaging, the photon statistics inherently follow a Poisson distribution, introducing uncertainty in the number of photons incident on the detector.

Consider imaging with a mean photon flux $\lambda$ per pixel. In a classical imaging scheme, the number of photons ($k$) detected in a pixel follows a Poisson distribution:
\begin{equation}
    P(k, \lambda) = \frac{\lambda^k e^{-\lambda}}{k!},
\end{equation}
where the standard deviation is $\sqrt{\lambda}$. This statistical uncertainty leads to variability in the measured transmission, defined as:
\begin{equation}
    t = \frac{N_{\text{measured}}}{N_{\text{incident}}},
\end{equation}
where $N_{\text{measured}}$ is the number of photons detected after the sample, and $N_{\text{incident}}$ is the number of photons incident on the sample. In classical imaging, $N_{\text{incident}}$ is approximated as the mean value $\lambda$ of the incident photon distribution. However, at low photon counts, the relative uncertainty in $N_{\text{incident}}$, given by $\frac{\sqrt{\lambda}}{\lambda}$, becomes significant. This high uncertainty degrades the signal-to-noise ratio (SNR) and necessitates using a larger photon dose to achieve reliable transmission measurements, as the limit of the uncertainty approaches 0 as $\lambda$ grows large.

Quantum imaging addresses this limitation in the photon sparse regime by utilizing the biphotons generated through spontaneous parametric down-conversion (SPDC). In this scheme, the strong number correlation between the paired photons allows direct measurement of $N_{\text{incident}}$ in the reference arm, significantly reducing the resulting uncertainty in $t$. This enables precise measurements, even in the low photon count regime, minimizing the dose delivered to the sample.

To perform such measurements, it is necessary to detect both coincident events—when paired photons are detected in the reference and test arms—and non-coincident events, which occur when a photon is absorbed or otherwise not transmitted through the sample. Accurate identification of SPDC photons versus background photons (e.g., diffusely scattered pump photons) is critical in this process. The ability to distinguish SPDC photons depends on the detector's energy resolution and the ratio of SPDC photons to background photons. For X-ray SPDC, achieving high energy resolution is challenging due to material limitations in current X-ray detector technologies, along with the large amount of background scattering produced in the non-linear diffraction process. Nevertheless, this capability is essential for effectively rejecting background events and achieving sub-shot-noise imaging in the X-ray regime.

\begin{figure}
\includegraphics[width=1.0\columnwidth]{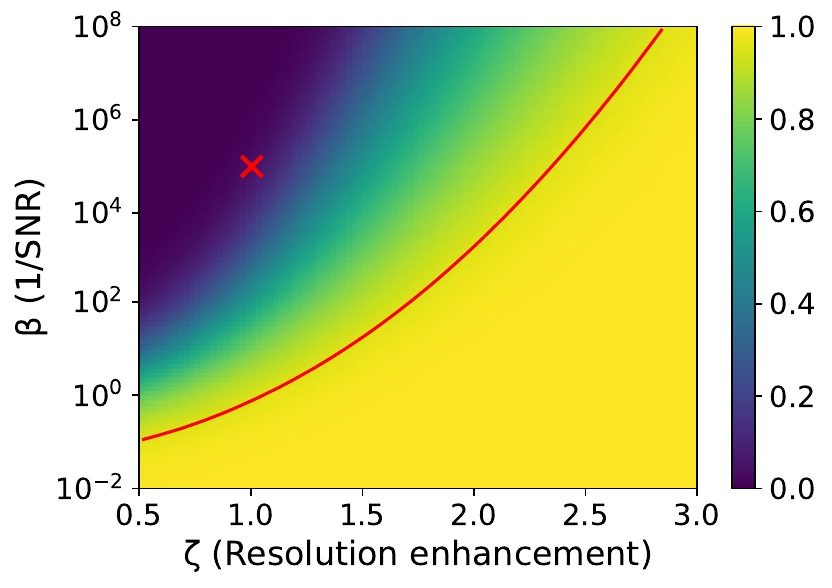}
\caption{SPDC Photon Identification Probability. The likelihood that an SPDC photon can be differentiated from a background photon as a function of the ratio of background events to down-converted events $\beta$ and the detector resolution enhancement $\zeta$. The red cross indicates the current measurements and the red contour line indicates where the likelihood of correct identification is 95\%. Realistic improvements along both axes will allow for high-likelihood identification of SPDC photons against background, enabling advanced quantum imaging modalities such as sub-shot-noise imaging.}
\label{fig6}
\end{figure}

A simple model using Bayesian probability can be used to determine the capability of the detection system to identify single SPDC photons, as a function of the energy resolution and the SNR of down-converted and background photons. In this model, the aggregate likelihood of correctly identifying an SPDC photon as such is given by:

\begin{equation}
    \mathcal{P}_{SPDC} = \int_{0}^{\infty} h(t) P(SPDC|t) dt
\end{equation}

where $h(t)$ is the detector ToT spectra of the down-converted photons and $P(SPDC|t)$ is the conditional probability a measured event with time-over-threshold $t$ is an SPDC photon, given by:

\begin{equation}
P(SPDC|t) = \frac{h(t)}{h(t) + \beta g_{\text{15 keV}}(t)}
\end{equation}

where $g_{\text{15 keV}}(t)$ is the ToT spectrum of the pump (15 keV) photons and $\beta$ is the ratio of background photons to SPDC photons on the detector. Both $h(t)$ and $g_{\text{15 keV}}(t)$ can also be functions of a resolution improvement factor $\zeta$, which impacts the standard deviations of ToT spectra (e.g., $\sigma_{ToT}' = \sigma_{ToT}/\zeta$). The functions $g_{\text{15 keV}}(t)$ and $h(t)$ can be numerically ascertained from data, such as that provided in Fig. \ref{fig2}(a). 

In our setup, the energy resolution is approximately 2 keV FWHM, with an SPDC-to-background photon ratio on the order of $\beta \approx 10^5$. Given these parameters and the pump photon energy of 15 keV, the numerical calculation of equation (9) indicates an aggregate probability of approximately 4$\%$. However, modest improvements—such as doubling the energy resolution ($\zeta = 2$) and reducing the background ratio by a couple orders of magnitude ($\beta \approx 10^3$)—could increase the identification probability of single SPDC photons to over 95\% (Fig. \ref{fig6}). Such enhancements appear feasible with advances in detector technology. For example, Timepix4 detectors, which are commonly known for their enhanced timing resolution, are also designed to achieve a twofold improvement in energy resolution ($<$1 keV FWHM) \cite{tpx4}. Additionally, the use of high-order crystalline reflections, such as (400) or (660) reflections at $2\theta$ angles approaching $90^\circ$, SPDC at the Brewster angle, and in Laue geometries, have been shown to significantly suppress background photons \cite{Shwartz,apsspdc,Borodin}. These improvements would enable precise SPDC photon identification, a critical step for realizing sub-shot-noise imaging. We anticipate that these measurements will be achievable and plan to target them in future studies.

\section{Conclusion}

Our experimental analysis at the CHX beamline of the NSLS-II facility at Brookhaven National Laboratory marks a significant advance in X-ray quantum imaging. Utilizing a pixelated detector, we successfully visualized the characteristic ring pattern typical of photon SPDC, and we measured spatial correlations of photon pairs at a rate of approximately $\sim6.3 \times 10^3$ pairs per hour. We utilized this source to perform coincidence imaging of several tests objects, including a biological sample (E. cardamomum seed pod). Our study uncovered how crystalline imperfections and pump divergence impact the properties of the down-converted properties and presented a framework for correcting the resulting aberrations in the imaging process, enhancing image clarity. Both simulated and digitized images were aligned with experimental outcomes, reinforcing the understanding of the theoretical and practical aspects of this process. Finally, we elucidated the path towards quantitative demonstration of quantum-enhanced transmission measurements through modest improvements in detector technology and background suppression in the down-conversion process.

This research sets the stage for innovative applications in quantum imaging, especially in fields where reducing radiation exposure is critical, such as in the study of sensitive biological materials. The record detection rates of X-rays via SPDC demonstrate the feasibility of employing quantum imaging techniques in areas previously limited by technological constraints. Our achievements in establishing quantum correlation imaging capabilities pave the way for the development of super-resolution imaging and diffraction methods, offering a novel approach to viewing complex structures with unprecedented detail and reduced dose. The measurements and simulations not only deepen our understanding of SPDC properties but also lay the groundwork for future advancements in optimizing quantum imaging systems. As we move forward, leveraging these insights will be essential in refining and revolutionizing approaches in both material and biological sciences at high-intensity X-ray sources such as synchrotrons and free electron lasers.

\begin{acknowledgments}

NSLS-II, a national user facility at Brookhaven National Laboratory (BNL), is supported in part by the U.S. Department of Energy, Office of Science, Office of Basic Energy Sciences Program under contract number DE-SC0012704. This project is supported through DOE-BER, Bioimaging Science Program KP1607020, within Biological Systems Science Division BSSD's Biomolecular Characterization and Imaging Science portfolio. It was also supported in part by the U.S. Department of Energy, Office of Workforce Development for Teachers and Scientists (WDTS) under the Science Undergraduate Laboratory Internships Program (SULI), and by the BNL Physics Department under the BNL Supplemental Undergraduate Research Program (SURP). This work was supported by the U.S. Department of Energy QuantISED award and BNL LDRD grants 19-30 and 22-22. 

We would like to recognize: Bryan Marino and Rick Greene, from NSLS-II's Complex Scattering Support team, for their assistance in fabricating our detector windows, beamstops, and their general support in maintaining beamline operations; John Lara, from NSLS-II's Structural Biology Support team, who made the crystal holders and the miniature goniometer support; Erik Muller, from the BNL Instrumentation Division, for loaning the diamond crystal used in our study; Erik Hogenbirk and Erik Maddox, from Amsterdam Scientific Instruments, for helpful discussions on time and ToT resolutions, and for providing driver updates for the Lynx T3 detector; Peter Schwander and Gabriel Beiner from University of Wisconsin Milwaukee and Andy Aquila, James Baxter, and Nick Hartley from Stanford's LCLS-II for helpful discussions; and Sylke Froechtenigt for help with Fig. \ref{fig1}(a).

\end{acknowledgments}

\appendix

\section{Appendixes}

\subsection*{Appendix 1: Phase Matching, Conservation of Momentum, and Emission Angles}
\setcounter{equation}{0} 
\renewcommand{\theequation}{A1.\arabic{equation}}

A pump of energy $E_{in}$ and momentum $\textbf{k}_{in}$ is aligned to a Bragg angle of $\theta$ to undergo diffraction via a crystallographic plane. The diffraction angle is detuned by $\Delta \theta$, which causes a deviation between $\textbf{k}_{in}$ and the reciprocal lattice vector $\textbf{G}$. This results in a phase-matching condition to instantiate X-ray SPDC of signal and idler photons.

Conservation of energy requires:

\begin{equation}
E_s = b E_{pump}
\end{equation}

and 

\begin{equation}
E_i = (1-b) E_{pump}
\end{equation}

where $b$ is a dimensionless constant $0 < b < 1$.

This results in:

\begin{equation}
|\textbf{k}_s| = b |\textbf{k}_{in}|
\end{equation}

and 

\begin{equation}
|\textbf{k}_i| = (1-b) |\textbf{k}_{in}|
\end{equation}

Conservation of momentum requires:

\begin{equation}
\textbf{k}_{out} = \textbf{k}_{in} + \textbf{G} = \textbf{k}_i + \textbf{k}_s  
\end{equation}
    
This results in the following equations of conservation of momentum:

\begin{equation}
|\textbf{k}_s| \sin(\alpha(b)) - |\textbf{k}_i| \sin(\alpha(1-b)) = 0
\end{equation}

and 

\begin{equation}
|\textbf{k}_s| \cos(\alpha(b)) + |\textbf{k}_i| \cos(\alpha(1-b)) = |\textbf{k}_{in} + \textbf{G}|
\end{equation}

where $\alpha(b)$ is the angle between $\textbf{k}_s$ and $\textbf{k}_{out}$ and $\alpha(1-b)$ is the angle between $\textbf{k}_i$ and $\textbf{k}_{out}$.

From the phase-matching condition:

\begin{equation}
|\textbf{k}_{in} + \textbf{G}| = |\textbf{k}_{in}| - |\textbf{k}_{in}| \Delta \theta \sin(2 \theta) = |\textbf{k}_{in}| (1 - \Delta \theta \sin(2 \theta))
\end{equation}

Substituting for $|\textbf{k}_{in} + \textbf{G}|, |\textbf{k}_s|,$  and $|\textbf{k}_i|$, the conversation of momentum equations become:

\begin{align}
b |\textbf{k}_{in}| \sin(\alpha(b)) - (1-b) |\textbf{k}_{in}| \sin(\alpha(1-b)) &= 0 \\
b \sin(\alpha(p)) - (1-b) \sin(\alpha(1-b)) &= 0
\end{align}

and

\begin{align}
b |\textbf{k}_{in}| \cos(\alpha(b)) + (1-b) |\textbf{k}_{in}| \cos(\alpha(1-b)) &= |\textbf{k}_{in}| (1 - \Delta \theta \sin(2 \theta)) \\ 
b \cos(\alpha(b)) + (1-b) \cos(\alpha(1-b)) &= c
\end{align}

where $c = 1 - \Delta \theta \sin(2 \theta)$.

In order to solve for $\alpha(b)$, a tactic is to use the binomial and small angle approximations. We note we can write:

\begin{equation}
\sin(\alpha(1-b)) = \frac{b}{1-b} \sin(\alpha(b))    
\end{equation}

and using $\cos(a) = \sqrt{1-\sin^2(a)}$:

\begin{equation}
\cos(\alpha(1-b)) = \sqrt{1 - \sin^2(\alpha(1-b))} = \sqrt{1 - \frac{b^2}{(1-b)^2} \sin^2(\alpha(b))}    
\end{equation}

Using the small angle approximations $\sin(a) \approx a$:

\begin{equation}
\cos(\alpha(1-b)) \approx \sqrt{1 - \frac{b^2}{(1-b)^2} \alpha(b)^2}    
\end{equation}

And then then binomial approximation $(1+a)^\frac{1}{2} \approx 1+\frac{a}{2}$:

\begin{equation}
\cos(\alpha(1-b)) \approx 1 - \frac{b^2}{2(1-b)^2} \alpha(b)^2    
\end{equation}

Similarly for $\cos(\alpha(b))$:

\begin{equation}
\cos(\alpha(b)) \approx 1 - \frac{1}{2} \alpha(b)^2    
\end{equation}

We can then substitute these into the second equation of conversation of momentum, equation (A1.12):

\begin{equation}
b \left(1 - \frac{1}{2} \alpha(b)^2 \right) + (1-b) \left(1 - \frac{b^2}{2(1-b)^2} \alpha(b)^2\right) \approx c
\end{equation}

Solving for $\alpha(b)$:

\begin{align}
b - \frac{b}{2} \alpha(b)^2 + (1 - b) - \frac{b^2}{2(1-b)} \alpha(b)^2 &\approx c \\ 
\left(- \frac{b}{2} - \frac{b^2}{2(1-b)}\right) \alpha(b)^2 &\approx c - 1 \\
\alpha(b) &\approx \sqrt{\frac{c - 1}{- \frac{b}{2} - \frac{b^2}{2(1-b)}}} \\
&\approx \sqrt{2(1-c)\frac{1-b}{b}} \\
&\approx \sqrt{2 \Delta \theta \sin(2 \theta) \frac{1-b}{b}}
\end{align}

Using $b = \frac{E_{SPDC}}{E_{pump}}$ and $\tan(\alpha) = \frac{r}{L}$, where r is the distance an SPDC photon from the Bragg center and L is the distance from diamond to the detector, this expression can be rewritten as:

\begin{equation}
    E_{SPDC}(r) \approx \frac{E_{pump}}{\frac{\arctan^2(\frac{r}{L})}{2 \Delta \theta sin(2 \theta)} + 1}
\end{equation}

which is equation (4) in the manuscript.

Here we present a new derivation which finds $\alpha(b)$ without the use of any approximations. 

We reorder the terms of the two conversation of momentum equations (A1.10 and A1.12) and square both sides. For A1.10:

\begin{align}
b \sin(\alpha(b)) &= (1-b) \sin(\alpha(1-b)) \\
b^2 \sin^2(\alpha(b)) &= (1-b)^2 \sin^2(\alpha(1-b))
\end{align}

and for A1.12:

\begin{align}
b \cos(\alpha(b)) &= c - (1-b) \cos(\alpha(1-b)) \\
b^2 \cos^2(\alpha(b)) &= (c - (1-b) \cos(\alpha(1-b)))^2 \\
 &= c^2 - 2c(1-b) \cos{\alpha(1-b)} + (1-b)^2 \cos^2(\alpha(1-b))
\end{align}

Sum the two equations (A1.26 and A1.29) and square the resulting equation. The LHS becomes:

\begin{align}
b^2 (\sin^2(\alpha(b)) + \cos^2(\alpha(b))) = b^2
\end{align}

And the RHS becomes:

\begin{align}
c^2 - 2c(1-b) \cos{\alpha(1-b)} + (1-b)^2 \left( \cos^2(\alpha(1-b)) + \sin^2(\alpha(1-b))\right) \\
= c^2 - 2c(1-b) \cos(\alpha(1-b)) + (1-b)^2
\end{align}

Thus, to find an expression for $\alpha(1-b)$:

\begin{align}
b^2 &= c^2 - 2c(1-b) \cos(\alpha(1-b)) + (1-b)^2 \\
\cos(\alpha(1-b)) &= \frac{b^2 - c^2 - (1-b)^2}{-2c(1-b)} = \frac{c^2 - 2b + 1}{2c(1-b)} \\
\alpha(1-b) &= \arccos\left(\frac{c^2 - 2b + 1}{2c(1-b)}\right) \\
&= \arccos\left(\frac{(1 -\Delta \theta \sin(2 \theta))^2 -2b + 1)}{2(1 - \Delta \theta \sin(2 \theta))(1-b)}\right)
\end{align}

And substituting $b=1-b$ to get an equation for $\alpha(b)$:

\begin{align}
\alpha(b) &= \arccos\left(\frac{c^2 + 2b - 1}{2cb}\right) \\
&= \arccos\left(\frac{(1 - \Delta \theta \sin(2 \theta))^2 + 2b - 1} {2(1 - \Delta \theta \sin(2 \theta)b}\right)
\end{align}

Equations (A1.23) and (A1.38) match very well so long as b is not very close to 0 or 1.

\setcounter{equation}{0} 
\renewcommand{\theequation}{A2.\arabic{equation}}

\subsection*{Appendix 2: Detuning Angle Calculations}

The detuning angle which produces a set of photons can be calculated from their emission angles by assuming conservation of energy. From equation (4) or (A1.24), an signal SPDC photon has energy:

\begin{equation}
E_s = \frac{E_{pump}}{\frac{\alpha_s^2}{2 \Delta \theta \sin(2 \theta)}+1}
\end{equation}

and its corresponding idler photon has energy:

\begin{equation}
E_i = \frac{E_{pump}}{\frac{\alpha_i^2}{2 \Delta \theta \sin(2 \theta)}+1}
\end{equation}

where $\alpha_s$ and $\alpha_i$ are the emission angles of the signal and idler photons, respectively. Letting $a = \frac{\alpha_s^2}{2 \sin(2 \theta)}$ and $b = \frac{\alpha_i^2}{2 \sin(2 \theta)}$ and using conservation of energy ($E_{pump} = E_s + E_i$), we can write:

\begin{align}
E_{pump} &= E_s + E_i = \frac{E_{pump}}{\frac{\alpha_1^2}{2 \Delta \theta \sin(2 \theta)}+1} + \frac{E_{pump}}{\frac{\alpha_2^2}{2 \Delta \theta \sin(2 \theta)}+1} \\
1 &= \frac{1}{\frac{a}{\Delta \theta}+1} + \frac{1}{\frac{b}{\Delta \theta}+1} 
\end{align}

Solving for $\Delta \theta$:

\begin{equation}
\Delta \theta = \sqrt{ab} = \frac{\alpha_s \alpha_i}{2 \sin(2 \theta)}
\end{equation}

\bibliography{references}

\end{document}